\documentclass[english,journal=apchd5]{achemso}
\usepackage[T1]{fontenc}
\usepackage[latin9]{inputenc}
\usepackage{units}
\usepackage{graphicx}
\PassOptionsToPackage{version=3}{mhchem}
\usepackage{mhchem}
\usepackage{subscript}
\usepackage{times}

\makeatletter

\title{Plasmon-assisted photoresponse in Ge-coated bowtie nanojunctions}
\author{Kenneth M. Evans}
\affiliation{Applied Physics Graduate Program}
\author{Pavlo Zolotavin}
\author{Douglas Natelson}
\affiliation[Department of Physics and Astronomy]{Department of Physics and Astronomy}
\alsoaffiliation[Department of Electrical and Computer Engineering]{Department of Electrical and Computer Engineering, Rice University,
Houston, Texas 77005}
\email{natelson@rice.edu}
\phone{(713) 348-3214}

  \usepackage[version=3]{mhchem}

\usepackage{babel}

\makeatother

\usepackage{babel}
\begin{document}
\begin{abstract}
We demonstrate plasmon-enhanced photoconduction in Au bowtie nanojunctions
containing nanogaps overlaid with an amorphous Ge film. The role of
plasmons in the production of nanogap photocurrent is verified by
studying the unusual polarization dependence of the photoresponse.
With increasing Ge thickness, the nanogap polarization of the photoresponse
rotates 90 degrees, indicating a change in the dominant relevant plasmon
mode, from the resonant transverse plasmon at low thicknesses to the
nonresonant ``lightning rod''mode at higher thicknesses. To understand
the plasmon response in the presence of the Ge overlayer and whether
the Ge degrades the Au plasmonic properties, we investigate the photothermal
response (from the temperature-dependent Au resistivity) in no-gap
nanowire structures, as a function of Ge film thickness and nanowire
geometry. The film thickness and geometry dependence are modeled using
a cross-sectional, finite element simulation. The no-gap structures
and the modeling confirm that the striking change in nanogap polarization
response results from redshifting of the resonant transverse mode,
rather than degradation in the Au/Ge properties. We note remaining
challenges in determining the precise mechanism of photocurrent production
in the nanogap structures.\bigskip{}

\end{abstract}


\bigskip{}
\bigskip{}

Light coupled into surface plasmons in metallic nanostructures can
generate extremely large enhancements to the local electromagnetic
field. Optical antennas exploit the resonant behavior and large scattering
cross section of plasmons in nanostructured metals for use in a wide
variety of applications including surface-enhanced Raman spectroscopy
(SERS)\cite{Ward2007,Ward2008}, subwavelength optics\cite{Bharadwaj2009,Neumann2013},
plasmonic optical trapping\cite{Zhang2010,Tsuboi2010,Righini2008,Righini2007},
and plasmon-assisted light harvesting\cite{Knight2011,Knight2013,Sobhani2013,Atwater2014,Li2014a,Kim2015,Collin2011,Balram2013,Chalabi2014}.
Photosensitive devices utilizing plasmonic nanoantennas generate photocurrent
through two dominant mechanisms: Near-field photons can be absorbed
in a nearby semiconductor to create electron-hole pairs; or high-energy
(``hot'') electrons generated by the plasmon can tunnel across a
semiconductor-metal potential barrier, directly resulting in photocurrent.
Plasmonically active devices often have the benefit of being highly
polarization dependent, operational at small biases, and capable of
being fabricated with photoactive elements on a scale smaller than
the wavelength of the incident light.

Gold ``bowtie'' nanojunctions, consisting of a nanowire (with or
without a nanogap) connected by two extended leads, have been shown
to act as combined optical antennas and electrical probes\cite{Natelson2013}.
The plasmon response in this geometry is dominated by a transverse
dipolar mode, resonant for incident light polarized perpendicular
to the length of the constriction. In devices with nanogaps at their
center, made through either electromigration or via a multi-step ``self-aligned''
lithographic technique\cite{Fursina2008}, the transverse plasmon
mode hybridizes with multipolar ``dark'' modes at the gap edges
to form a strong electromagnetic field enhancement at the junction
center, or ``hotspot''. The SERS response of molecules assembled
into these hotspots follows the signature polarization dependence
of dipolar optical antennas, with maximum response perpendicular to
the axis of the constriction\cite{Herzog2013}. This is in contrast
to expectations of the nonresonant ``lightning rod'' effect, when
the maximum plasmon response occurs when the incident polarization
parallel to the elongated direction of a metal wire or tip, from the
excitation of longitudinal or tip plasmons\cite{NanoOptics2006}.

In no-gap bowtie nanowires, resonant laser illumination produces a
polarization-dependent hotspot via photothermally induced changes
in resistivity\cite{Herzog2014}. Direct absorption and the excitation
of the transverse plasmon mode locally increase the temperature of
the nanowire, decreasing its conductance, $-\Delta G$, due to the
temperature-dependent resistivity of the metal. The polarization dependence
of this change is also dipolar, so that the temperature of the nanowire
increases more for the transverse excitation, even without the existence
of gap plasmons. This measurement provides a convenient measurement
technique to assess the plasmon response of the nanowire material
itself.

In this report, we study the photoresponse of bowtie nanoantennas
overlaid with a thin Ge film. We examine the plasmon-assisted photoconduction
($I_{photo}$) in bowties with nanogaps. At low Ge thicknesses, we
find that the photoconductance is maximal with incident polarization
transverse to the nanowire; in contrast at higher Ge thicknesses,
the maximum photoconductance occurs when the incident polarization
is rotated by 90 degrees. The Ge thickness alters which plasmon response
is dominant in the nanogaps, the resonant transverse mode (at low
thicknesses) or the nonresonant ``lightning rod'' response (at high
thicknesses). A concern is whether the transverse plasmon mode is
eliminated entirely due to degradation of the Au properties with thicker
Ge layers. To test for this, we also measured the plasmon-induced
heating in unbroken (no-gap) nanowires ($-\Delta G$). We find experimentally,
and in accord with a computational model, that the presence of Ge
predictably redshifts the resonance of the nanoantenna's transverse
plasmon mode without degrading the Au plasmonic response. The manipulation
of resonant and nonresonant plasmon responses is an interesting avenue
to consider in further plasmonic optoelectronic structures.

This work is one of a handful demonstrating plasmon assisted photocurrent
in individual nanoantennas. Unlike similar work integrating photosensitive
materials, such as graphene\cite{Shi2011,Gabor2011}, MoS\textsubscript{2}\cite{Hong2015},
and traditional semiconductors\cite{Tang2008,Cao2009,Falk2009,Mousavi2014},
into plasmon resonant geometries, the nanogap devices reported in
this study rely on a truly nanoscale active optical region, typically
100 $\unit{nm}$ $\times$ 5 $\unit{nm}$ $\times$ 10 $\unit{nm}$,
fabricated from amorphous Ge, an abundant and easily deposited material.
Other schemes for fabricating field enhanced photodetectors and photovoltaic
devices have been discussed in depth over the past five years\cite{Bharadwaj2009,Atwater2010,Clavero2014,Brongersma2015}.
This report builds upon recent work outlining the strength of non-radiative,
hybridized plasmon modes in bowtie nanoantennas.

\section*{Methods}

The devices in this report are fabricated on n-type Si wafers with
a 200 $\unit{nm}$ thermally grown oxide layer. Electron beam lithography
is used to define the nanowire constriction and small triangular electrodes,
which extend to larger, prefabricated Au contact pads deposited using
a shadowmask. After development, a 1 $\unit{nm}$ Ti adhesion layer
and 13 $\unit{nm}$ of Au are deposited by electron beam evaporation,
followed by lift-off in acetone. Nanowires are patterned to be 600
$\unit{nm}$ long and vary from 80 $\unit{nm}$ to 140 $\unit{nm}$
wide depending on the desired geometry.

\begin{figure}
\includegraphics[scale=0.95]{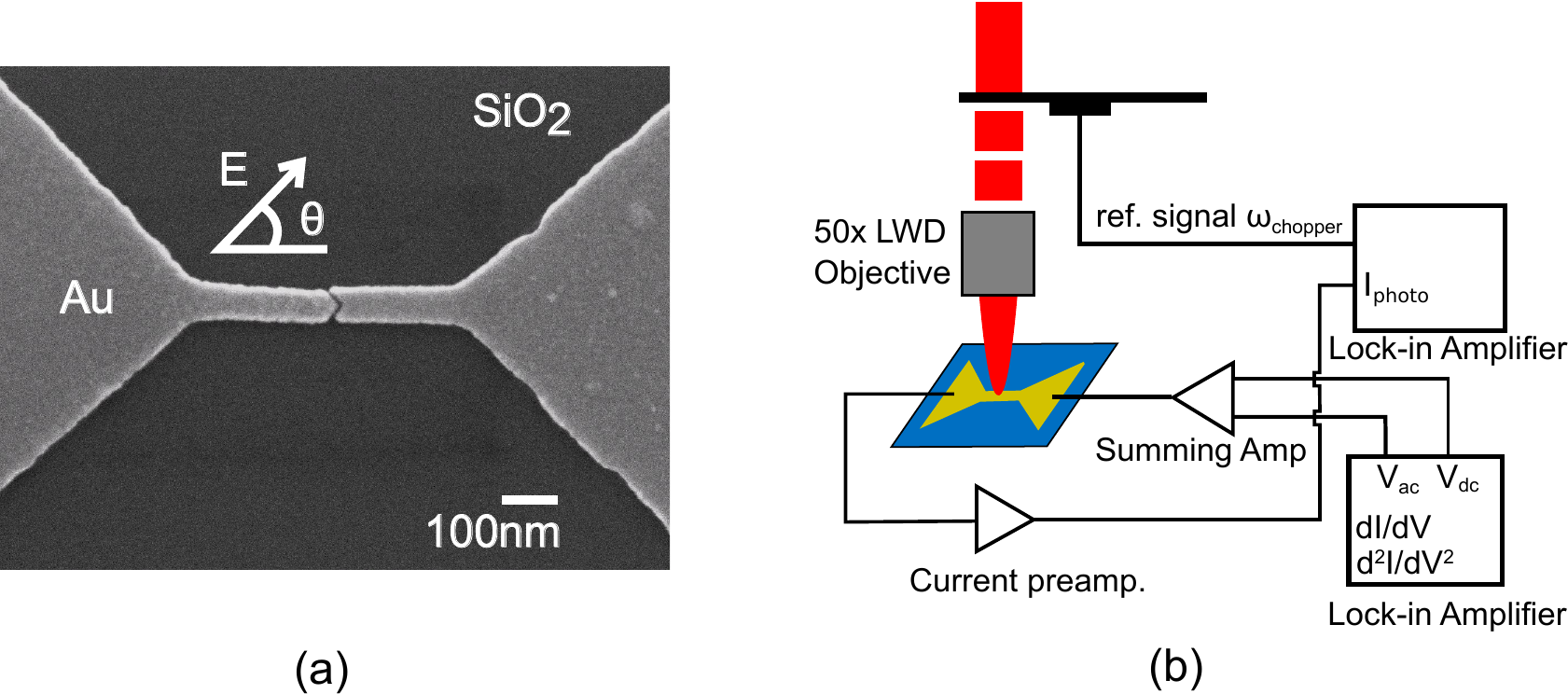}

\protect\protect\caption{\label{fig:figschem}(a) Scanning electron microscope (SEM) image
of a typical self-aligned device with 80 +/- 5 $\unit{nm}$ width
coated in a 10 $\unit{nm}$ Ge film. (b) A schematic diagram of the
experimental setup. Devices are biased with a dc voltage summed with
a small (typically 10 $\unit{mV}$) sinusoidal ac excitation voltage.
Signal passes through a current preamplifier to two separate lock-in
amplifiers, one synched to the ac frequency to measure the first and
second harmonic of the differential conductance and a second synched
to an optical chopper to measure photoresponse.}
\end{figure}

In the case of ``self-aligned'' nanojunctions, an extra step of
lithography is needed to create the gap\cite{Fursina2008}. In the
first step, the left half of the nanowire is patterned and the 1 $\unit{nm}$
/ 13 $\unit{nm}$Ti/Au combination is deposited, followed by a 1 $\unit{nm}$
layer of \ce{SiO2} and 14 $\unit{nm}$ layer of Cr. After the metal
deposition, exposing the sample to air creates a layer of Cr-oxide
which swells to extend just past the edge of the pattern, acting as
a shadowmask in the second evaporation. The layer of \ce{SiO2} prevents
Cr and unwanted elements of the Cr etchant (namely cerium) from contaminating
the Au and potentially degrading device plasmon response. During a
second lithographic step, the right half of the pattern is aligned
to the existing left half. An identical evaporation follows with the
same materials and thicknesses, to ensure device uniformity. The Cr
and \ce{SiO2} are etched away, leaving a 2-10 $\unit{nm}$ gap between
the two sides of the nanowire where the oxidized Cr expanded beyond
the first pattern (\ref{fig:figschem}a). The width of the gap can
be tailored by adjusting the thickness of the Cr layer\cite{Fursina2008}.

The devices are cleaned with \ce{O2} plasma and wirebonded to a chip
carrier. Each device is scanned with a 785 $\unit{nm}$ diode laser
configured with a telescopic lens rastering system to create a spatial
map of the photoresponse. The laser has a Gaussian beam profile with
a full width at half-maximum of approximately 1.8 $\unit{\mu m}$,
operating in CW mode at a typical laser power of 3-4 $\unit{mW}$
reaching the sample. A lock-in amplifier, synched to an optical chopper
at a frequency of 281 $\unit{Hz}$, is used to measure device optical
response: either the photocurrent generated in self-aligned devices
($I_{photo}$) or the optically induced change to the conductance
in no-gap nanowires ($\Delta G=\nicefrac{\Delta I}{V_{dc}}$). The
sample is biased with the summed output of the desired dc voltage,
typically up to 0.2 $\unit{V}$, and a 10 $\unit{mV}$ RMS ac voltage,
and a second lock-in, at frequency 789 $\unit{Hz}$, measures the
first- and second-order differential conductance ($\nicefrac{dI}{dV}$
and $\nicefrac{d^{2}I}{d^{2}V}$). The dc current ($I_{dc}$) is measured
using a digital-to-analog converter (DAC) built into one of the lock-in
amplifiers (\ref{fig:figschem}b). After each scan, the laser is then
finely positioned at the hotspot center by maximizing $I_{photo}$
or $-\Delta G$ , the polarization is rotated continuously from $\theta=0\textdegree$
to $\theta=360\textdegree$ (with $\theta=0\textdegree$ defined to
be along the nanoantenna long axis) using a motorized half-wave plate,
and the optical and electrical responses are measured. Bare bowtie
nanowires (with and without nanogaps) are coated with an amorphous
Ge film of varying thickness deposited by electron beam evaporation,
and remeasured using the same procedure. All measurements presented
in this paper were conducted at room temperature at a pressure on
the order of 100 $\unit{mTorr}$.

\section*{Results}

\begin{figure}
\includegraphics[scale=0.7]{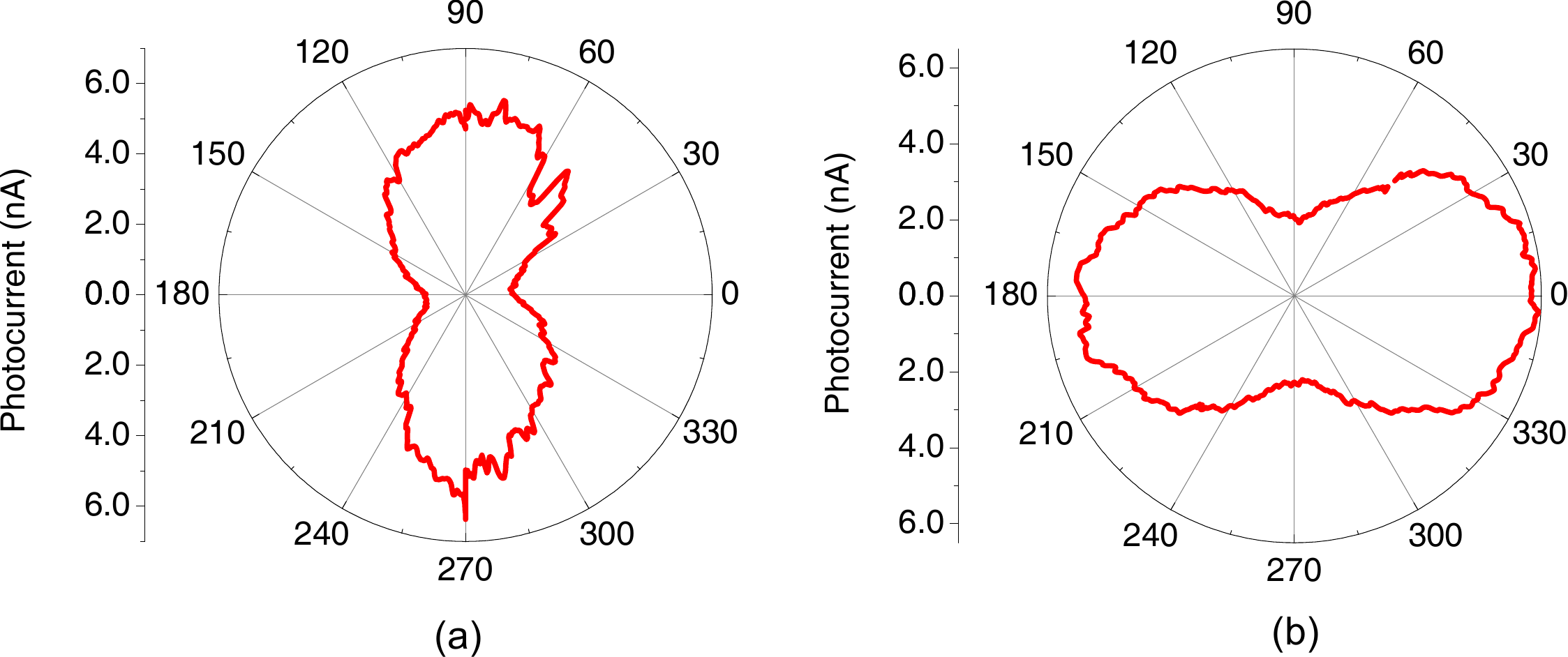}\protect\protect\caption{\label{fig:pc}(a) The dipolar polarization dependence of an 80 $\unit{nm}$
wide nanoantenna with a 10 $\unit{nm}$ Ge film where the transverse
plasmon mode is roughly on resonance with the 785 $\unit{nm}$ laser.
(b) The dipolar polarization of an 130 $\unit{nm}$ wide nanoantenna
with a 35 $\unit{nm}$ Ge film where the ``lightning rod'' mode
is roughly on resonance with the 785 $\unit{nm}$ laser. }
\end{figure}

Photoresponse of the self-aligned nanogap junctions is dominated by
a positive photoconductance of the Ge in the gap region. Photocurrent
measurements of self-aligned nanojunctions present two distinct polarization
dependences. In devices with a nanowire width of 80 $\unit{nm}$ and
a 10 $\unit{nm}$ Ge film, the photocurrent follows the signature
dependence of dipolar optical antennas, with maximum response at $\theta=90\textdegree$
(\ref{fig:pc}a). Repeating this experiment with 130 $\unit{nm}$
wide nanowire and a thicker layer of Ge (35 $\unit{nm}$) shifts the
polarization dependence by $90\textdegree$(\ref{fig:pc}b). In both
cases, photocurrent is linear with respect to laser power and dc bias,
and only occurs at the junction center.

To understand this, we consider the plasmons in the system. In previous
work exploring photoresponse in bowtie nanoantennas, it has been demonstrated
that the greatest field enhancement occurs when incident light is
polarized perpendicular to the nanowire axis, corresponding to a strong
dipolar transverse plasmon mode\cite{Herzog2013}. For bare Au nanowires,
this mode is known to be resonant at 785 $\unit{nm}$ for bare nanowires
\textasciitilde{}130 $\unit{nm}$ wide and 13 $\unit{nm}$ thick.
It would appear that in the narrower-wire nanogap devices with the
thinner Ge film, the photocurrent is dominated by this transverse
plasmon response, while for the wider-wire nanogap devices with thicker
Ge the dominant plasmon response is that of the nonresonant longitudinal
``lightning rod'' mode.

\begin{figure}
\includegraphics[scale=0.9]{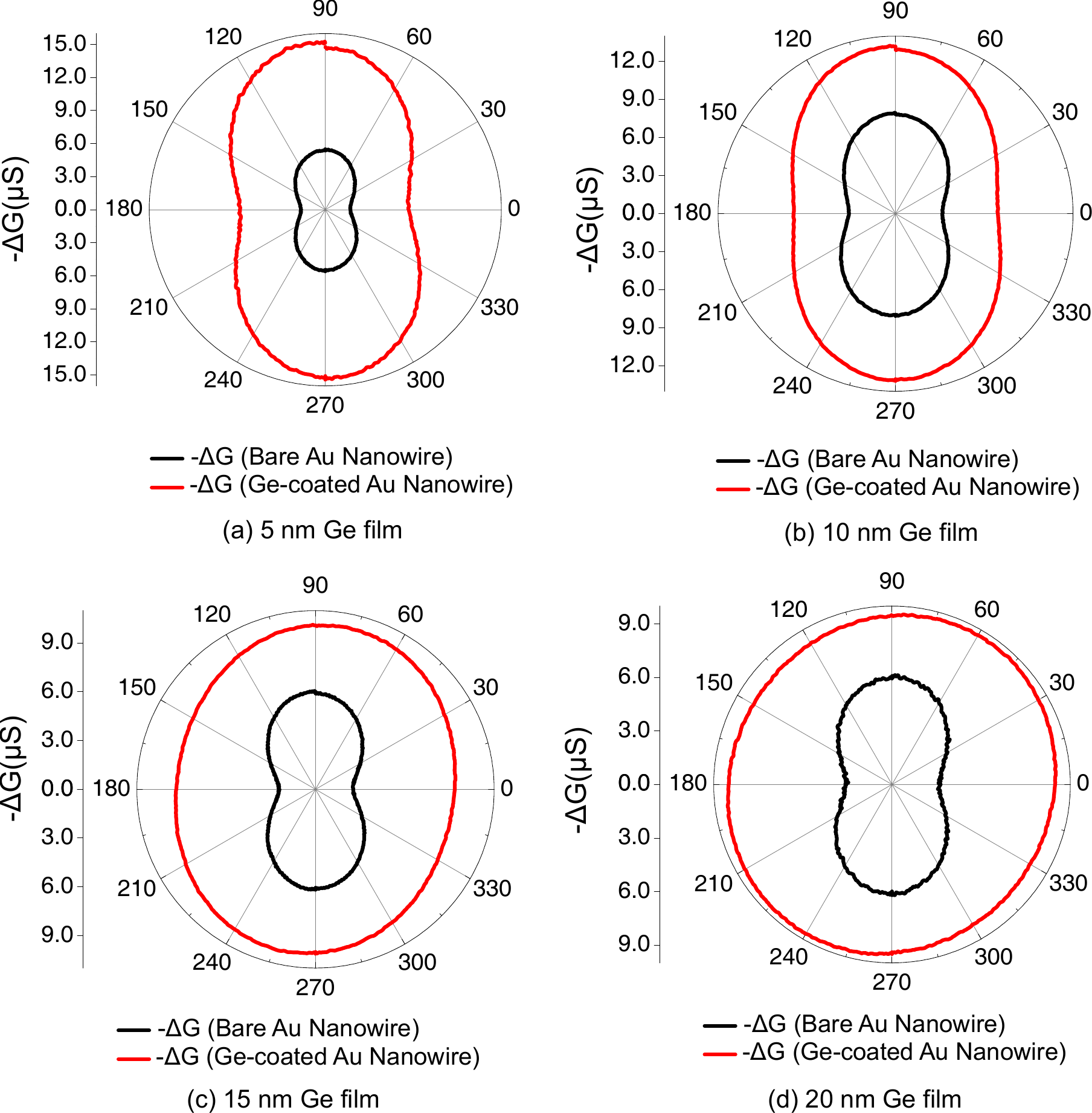}\protect\protect\caption{\label{fig:thicknesses}Experimental data displaying the change in
polarization dependence for ``unbroken'' nanowires with varying
layers of Ge deposited on top. Each successive trial becomes more
circular as the resonance of the transverse mode moves further away
from the laser wavelength.}
\end{figure}

An issue is whether, in the thicker Ge devices, the plasmonic properties
of the Au itself is degraded (e.g., through partial alloying with
the Ge due to the longer deposition), or whether the change in the
plasmon response is due purely to shifting of the resonant transverse
mode away from the operating wavelength. To understand this dramatic
change in nanogap device plasmon response, we fabricate ``unbroken''
no-gap nanowires, and assess the plasmonic properties of the Au through
study the photothermally induced change in the conductance, $-\Delta G$
(due to the temperature-dependent Au resistivity), as a function of
nanowire width and Ge film thickness. Measurements on bare Au bowtie
nanowires (\textasciitilde{}130 $\unit{nm}$ wide and 13 $\unit{nm}$
thick) are consistent with previous work measuring resistive heating
in these structures; the conductance change includes a characteristic
$cos^{2}\left(\theta\right)$ polarization dependent contribution,
with a maximum at $\theta=90\textdegree$ due to the contribution
to heating from the structure's transverse plasmon mode. \cite{Herzog2014}.
This indicates that the laser wavelength is on resonance with the
transverse mode of the nanowire (width \textasciitilde{}130 $\unit{nm}$,
thickness 13 $\unit{nm}$), heating it approximately by 2-10$\unit{K}$
depending on the exact width of the device and the incident laser
power. The increase in temperature can be estimated directly from
the change in conductance using the formula provided in Herzog, et.
al. (2014), $\triangle T=\frac{-R^{2}\triangle G}{\frac{dR}{dT}}$,
where $R$ is the resistance of the device and $\frac{dR}{dT}=0.12$
for these Au bowtie nanowires. This value has been measured to be
consistent for many nanowire devices.

\begin{figure}
\includegraphics[scale=0.9]{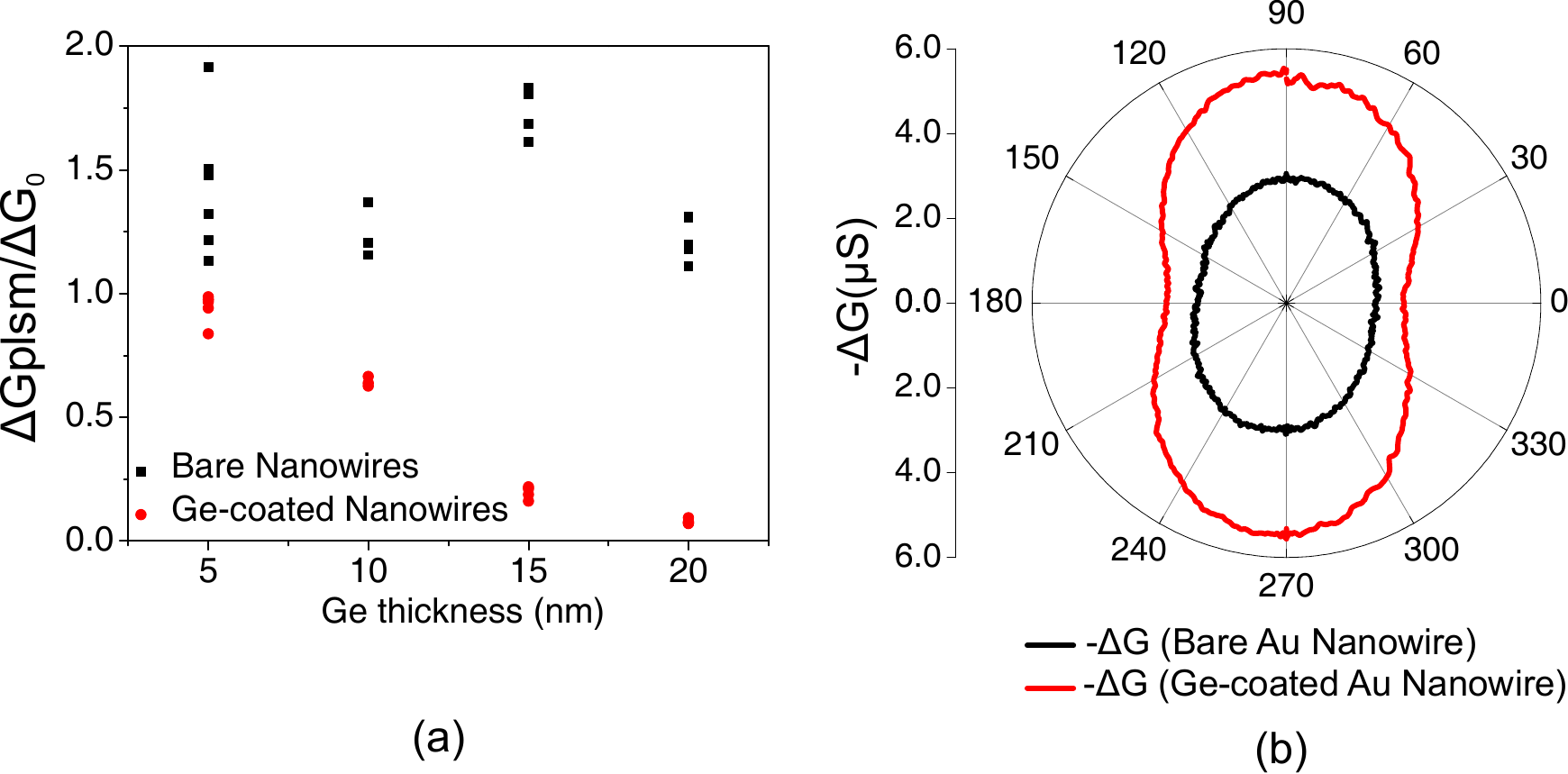}

\protect\protect\caption{\label{fig:ratiothin}(a) Each point represents the $\nicefrac{\Delta G_{plsm}}{\Delta G_{0}}$
ratio measured for an individual device, before and after the Ge film
deposition. The values for bare nanowires (black squares, with each
batch of bare wires marked at the Ge thickness eventually coated)
are roughly all the same for all measured bare junctions, which vary
in width from device to device. The values for Ge-coated nanowires
(red circles) have decreasing $\nicefrac{\Delta G_{plsm}}{\Delta G_{0}}$
ratios for increasing Ge thickness, indicating that their polar plots
are more circular with less contribution of the transverse plasmon
mode toward the total photoresponse. (b) A polar plot of an off-resonance,
bare device moved closer to resonance with the addition of a 10 $\unit{nm}$
Ge film. In this case the $\nicefrac{\Delta G_{plsm}}{\Delta G_{0}}$
ratio increases.}
\end{figure}

After each initial characterization of bare devices, a Ge film of
varying thickness (5 $\unit{nm}$, 10 $\unit{nm}$, 15 $\unit{nm}$,
and 20$\unit{nm}$) is deposited by electron beam evaporation and
the photoresponse is remeasured (\ref{fig:thicknesses}). For each
trial, the polar plots become progressively more circular, as the
increasing thickness of Ge redshifts the resonance of the transverse
plasmon mode away from the laser wavelength of 785 $\unit{nm}$. The
data are fitted to the expression $-\Delta G\left(\theta\right)=\Delta G_{0}+\Delta G_{plsm}cos^{2}\left(\theta\right)$.
Here, $\Delta G_{0}$ is the nonresonant, direct absorption contribution,
while $\Delta G_{plsm}$ is the component of the conductance change
due to dipolar plasmon-based heating. The ratio $\nicefrac{\Delta G_{plsm}}{\Delta G_{0}}$
then gives a measure of the transverse plasmon contribution to the
change in conductance for each thickness trial. This ratio is roughly
constant for bare junctions across all four trials, but steadily decreases
in the presence of increasing Ge film thickness (\ref{fig:ratiothin}a).
By 20 $\unit{nm}$ of Ge, the devices lose all their polarization
dependence; their polar plots become entirely circular, indicating
that the presence of the film has shifted the transverse plasmon mode
far off-resonance, removing the dipolar response of the nanoantenna,
and leaving behind only heating from nonresonant absorption.

To verify that the dominant effect on the polarization is redshifting
of the plasmon resonance rather than simple attenuation of the plasmon
resonance due to Au degradation, we fabricated intentionally off-resonance
nanowires, less than 100 $\unit{nm}$ wide, and then deposited a Ge
film to redshift the resonance back to the laser wavelength. The pre-
and post-Ge deposition bowties are analyzed following the above procedures
(\ref{fig:ratiothin}b). In this case we see the $\nicefrac{\Delta G_{plsm}}{\Delta G_{0}}$
ratio increase, indicating that the polarization dependence of the
junction becomes more dipolar in the presence of the Ge film. This
effect was found to be consistent in 8 measured devices.

\begin{figure}
\includegraphics[scale=0.95]{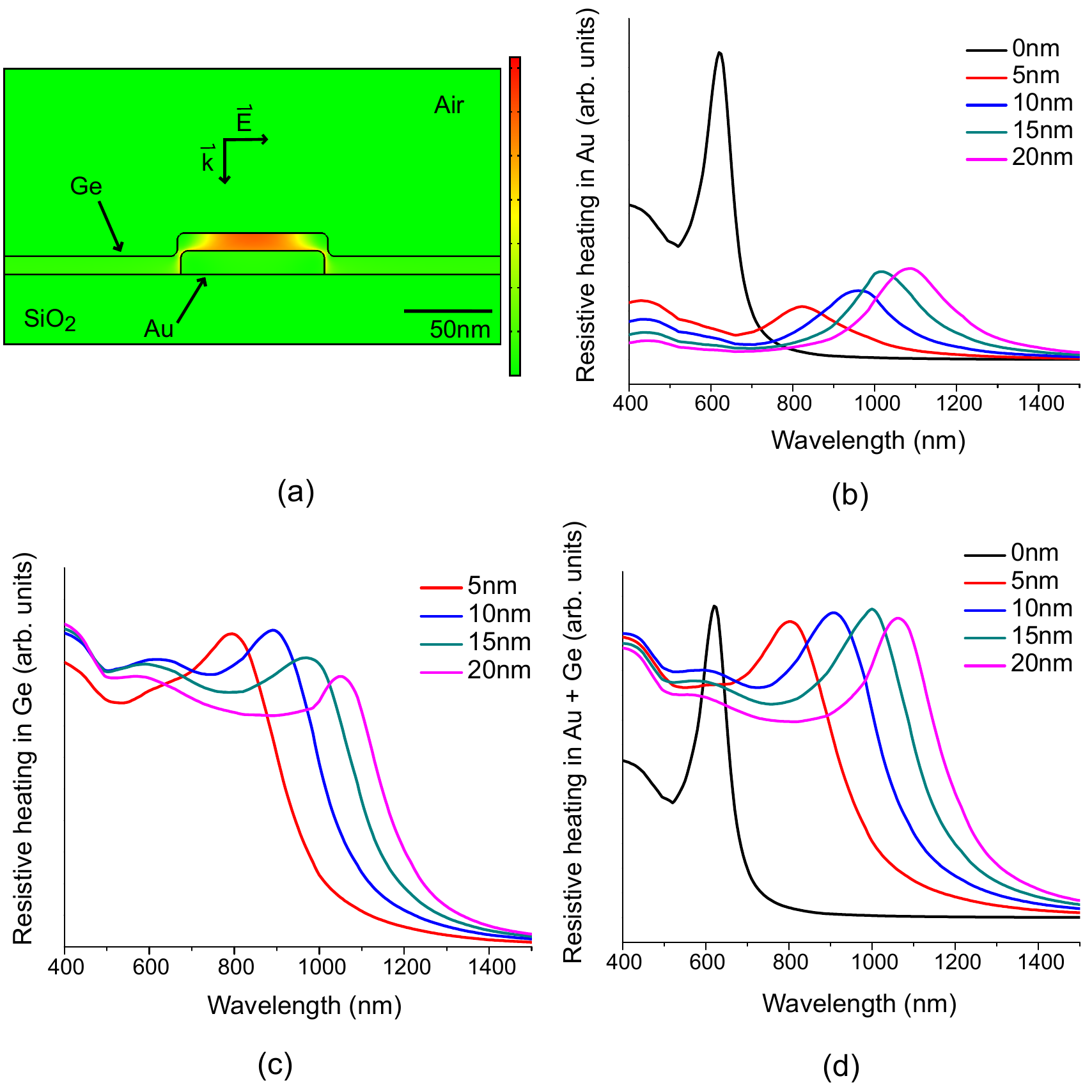}
\protect\protect\caption{\label{fig:comsol}(a) An image of the solution for the variable $Q$
plotted in arbitrary units, green indicating zero heating. (b) Spectra
of the variable $Q$ for an 80 $\unit{nm}$ nanowire in the 2D COMSOL
model for varying thicknesses of Ge. (c) Spectra of the variable $Q$
integrated over a portion of the Ge for an 10 $\unit{nm}$ film on
top of an 80 $\unit{nm}$ nanowire. (d) Spectra of the sum of the
resistive heating in the Au nanowire and the Ge film.}
\end{figure}

These experiments verify that for the gapped nanojunctions a 35 $\unit{nm}$
Ge film will remove the transverse plasmon mode's contribution toward
the photocurrent for a device previously on resonance with the laser
wavelength, as seen in \ref{fig:pc}b. For structures made off resonance
with the laser, adding an appropriately thick layer of Ge can tune
the device back onto the laser wavelength, as seen in \ref{fig:pc}a.
To map out the appropriate thicknesses of Ge over a range of device
widths, the resonance of the transverse plasmon mode in nanoantennas
was modeled with a finite-element software package designed to solve
differential equations over physical spaces (FEM, COMSOL 3.5a). A
cross-sectional model was designed so that the incident light is polarized
perpendicular (transverse magnetic mode) to an infinitely long nanowire,
isolating the transverse plasmon mode for the purposes of this study
(\ref{fig:comsol}a). Ge is known to be entirely amorphous when deposited
by evaporation with the substrate at room temperature, and accordingly
a dielectric function for amorphous Ge was used in the model\cite{Properties1970,Tomlin1976}.
The dielectric function for Au was taken from Johnson and Christy\cite{jc1972}.
Maxwell's equations are solved over the entire solution space as a
function of nanowire width and Ge film thickness. For an incident
plane wave with free space wavelength ranging from 400 $\unit{nm}$
to 1500 $\unit{nm}$, a scalar proportional to the optically driven
resistive heating is found by integrating resistive heating $Q=\frac{1}{2}Re\left(\sigma\left\Vert \overrightarrow{E}\right\Vert ^{2}\right)$
over the area of the Au, where $\sigma$ is the electrical conductivity
and $\left\Vert \overrightarrow{E}\right\Vert $ is the norm of the
total electric field (\ref{fig:comsol}b). As the film thickness increases,
the original resonance is redshifted and broadened, as expected from
previous work exploring plasmon modes in the presence of semiconductors\cite{Knight2009,Knight2013,Liu2011}.
Bare devices with transverse modes previously resonant at the laser
wavelength of 785 $\unit{nm}$ will be shifted off resonance further
into the near-IR in the presence of a 5-10 $\unit{nm}$ film of Ge.

This simple model provides a proof-of-concept explanation for the
shift in the resonance of the transverse plasmon mode, but does not
take into account heating from absorption in the Ge layer; it assumes
that all photothermally induced change in current comes from heating
in the Au alone. This calculational omission accounts for the predicted
decrease in the magnitude of the peaks for increasing Ge thickness
not seen in the experiment. In contrast, in the measurements the photothermal
heating in unbroken bowtie nanowires is larger with the Ge film than
without it (ie. $\Delta G_{0}$ of the Au is always larger for Ge-coated
nanowires than for bare nanowires). Integrating the variable $Q$
over the volume of the Ge film produces a broad plateau at a wavelength
below the plasmon resonance due to direct absorption in the Ge overlayer
(\ref{fig:comsol}c). Adding the two spectra for $Q$ together (ie.
\ref{fig:comsol}b + \ref{fig:comsol}c) results in resonance peaks
of Ge-coated nanowires with magnitudes for the structure's total optically
driven resistive heating comparable to the magnitude of bare junctions
(\ref{fig:comsol}d). Extracting precise predictions for the Au temperature
increase in Ge-coated nanowires under laser illumination would require
knowledge of both the thermal boundary resistance at the interface
between the Ge at the Au, and the thermal conductance of the amorphous
Ge.

\section*{Discussion}

We have found that the change in polarization response of the photoconductive
nanogap antennas with Ge thickness is caused by the tuning of which
plasmons dominate at the incident wavelength, the resonant transverse
mode or the nonresonant ``lightning rod'' response. In either limit
we have not addressed the precise mechanism of the positive photoconductance
of the Ge-coated nanogap structures. There are several potential mechanisms
which could contribute to the photocurrent generated in the Ge-coated
nanogap structures: direct absorption of light in the Ge located at
the gap creating electron-hole pairs via the photovoltaic effect (PVE);
driving ``hot'' carriers generated in the plasmon across the semiconductor-metal
barrier, also known as direct electron transfer (DET); or through
plasmon-induced resonant energy transfer (PIRET), where the plasmon
can directly excite electron-hole pairs in the Ge\cite{Cushing2012,Li2015a}.
Similar studies have also reported photocurrent generation due to
the photothermoelectric effect (PTE)\cite{Echtermeyer2014,Hong2015}
and optical rectification\cite{Shi2011,Rectification2010}. In this
study, photocurrent generation only occurs at the nanojunction center,
not along the length of the nanowire, and therefore cannot be due
to PTE. Additionally, the photocurrent does not trace the signature
nonlinear electrical response for optical rectification, $I_{photo}=\frac{1}{4}\frac{d^{2}I}{dV^{2}}V_{ac}^{2}$.
Therefore the photocarriers in Ge-coated bowtie nanojunctions must
be generated by a combination of field-enhanced PVE in the Ge and
by plasmonic coupling between the Au and the Ge through DET or PIRET.
DET has been reported in various plasmon-based photovoltaic devices,
but device responsivity is typically inefficient because collection
of hot carriers requires careful engineering of geometry and materials
to ensure effective coupling between the plasmon, absorption medium,
and electrodes\cite{Chalabi2014,Sobhani2013,Brongersma2015,Atwater2010}.
The measured responsivity in theGe-coated nanogaps in this works calculated
to be \textasciitilde{}10\textsuperscript{-6} $\unitfrac{A}{W}$
at 0.2 $\unit{V}$ which is comparable to the values listed in Shi,
et. al. (2011), who use electromigrated bowtie nanoantennas on top
of graphene, but is significantly lower than the values reported by
Tang, et. al. (2008). However, their device has a much larger active
area and uses crystalline Ge. All three potential mechanisms would
have identical polarization dependences; higher fields would produce
larger enhancements and more hot carriers, making it difficult to
distinguish between the these mechanisms based on photoresponse measurements
alone. Spectral data or time-resolved measurements could shed light
on the relative contribution of each process due to the characteristically
short lifetime of hot electrons.

Bowtie nanowires provide a platform for understanding plasmon response
in individual metal-semiconductor-metal nanoantennas. We provide experimental
and computational evidence that the dominant plasmon modes in such
structures may be controlled by tuning between resonant and nonresonant
responses. Photocurrent measurements of Ge-coated nanogap structures
dominated by the resonant transverse mode are consistent with measurements
of unbroken bowtie nanowires; polarization dependence in both cases
follows the known $cos^{2}\left(\theta\right)$ polarization dependence
of dipolar optical antennas, with peak photoresponse occurring at
an excitation perpendicular to the length of the nanowire ($\theta=90\textdegree$).
By optimizing device geometry, this resonant response may be ``tuned
away'' without degrading the Au properties, leaving behind the nonresonant,
longitudinal ``lightning rod'' mode. The nanogap devices operate
with a truly nanoscale active region defined by the volume of the
interelectrode nanogap. Knowledge of the impact of dielectrics on
both resonant and nonresonant local plasmon-enhanced fields is an
important step toward fabricating efficient nanoscale light harvesting
devices.
\clearpage

\begin{acknowledgement}
K.M.E. would like to acknowledge Robert A. Welch Foundation award
C-1636 and thank Will Hardy and Yajing Li for useful discussions during
the completion of this project. P.Z. and D.N. acknowledge support
from ARO award W911-NF-13-0476. 
\end{acknowledgement}

\begin{suppinfo}
\end{suppinfo}


\providecommand{\latin}[1]{#1}
\providecommand*\mcitethebibliography{\thebibliography}
\csname @ifundefined\endcsname{endmcitethebibliography}
  {\let\endmcitethebibliography\endthebibliography}{}

\begin{tocentry}
\includegraphics{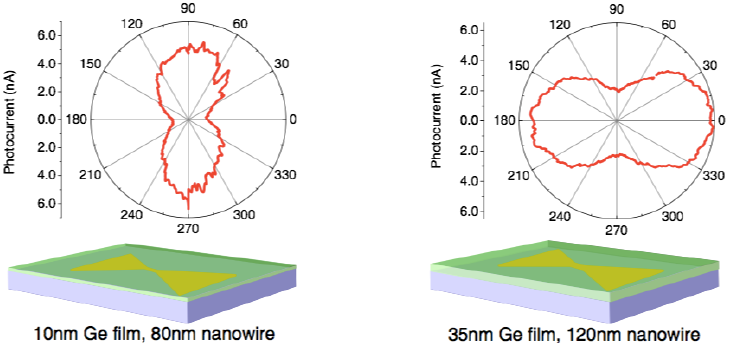}
\vspace{1cm}
  
For Table of Contents Use Only,  ``Plasmon-assisted photoresponse in Ge-coated
  bowtie nanojunctions'', by Kenneth M. Evans, Pavlo Zolotavin, and
  Douglas Natelson.
\end{tocentry}

\end{document}